\documentclass[twocolumn,twoside,slac_two]{revtex4}
\usepackage{graphicx}
\usepackage{fancyhdr}
\begin{document}

\title{Gravitational Lensing of Distant Supernovae}

%

\author{P. Premadi}
\affiliation{Bandung Institute of Technology, Bandung, Indonesia}
\author{H. Martel}
\affiliation{Universit\'e Laval, Qu\'ebec, Qc, G1K 7P4, Canada}

\begin{abstract}
We use a series of ray-tracing experiments to determine the
magnification distribution of high-redshift sources by gravitational
lensing. We determine empirically the relation between magnification
and redshift, for various cosmological models. We then use this
relation to estimate the effect of lensing on the determination of the
cosmological parameters from observations of high-$z$ supernovae. We
found that, for supernovae at redshifts $z<1.8$, the effect of lensing
is negligible compared to the intrinsic uncertainty in the
measurements. Using mock data in the range $1.8<z<8$, we show that the
effect of lensing can become significant. Hence, if a population of
very-high-$z$ supernovae was ever discovered, it would be crucial to fully
understand the effect of lensing, before these SNe could be used to
constrain cosmological models.
\end{abstract}

\maketitle

\thispagestyle{fancy}


\section{INTRODUCTION}
High-redshift supernovae have become a major tool in modern
cosmology. By measuring their apparent magnitudes, we can
estimate their luminosity distances $d_L$ 
(see \cite{tonryetal03,barrisetal04,riessetal04}, and references therein).
Since the relationship
between $d_L$ and the redshift $z$ depends on the cosmological parameters,
observations of distant SNe can constrain the cosmological model. Prior
to the announcement of the WMAP results \cite{bennettetal03}, 
observations of high-$z$ SNe provided the most compelling evidence of the
existence of a nonzero cosmological constant.

The luminosity distances $d_L$ are determined by combining the observed 
fluxes $F$ with estimates of the SNe luminosities $L$. Uncertainties 
in $d_L$ are caused by uncertainties in $L$, because
SNe are not perfect standard candles. The flux $F$ is much easier to measure,
but for distant sources the value of $F$ might be altered by gravitational
lensing cause by the intervening distribution of matter. For instance,
a positive magnification would result in a increase in $F$, and an 
underestimation of $d_L$.

\begin{figure*}[t]
\centering
\includegraphics[width=145mm]{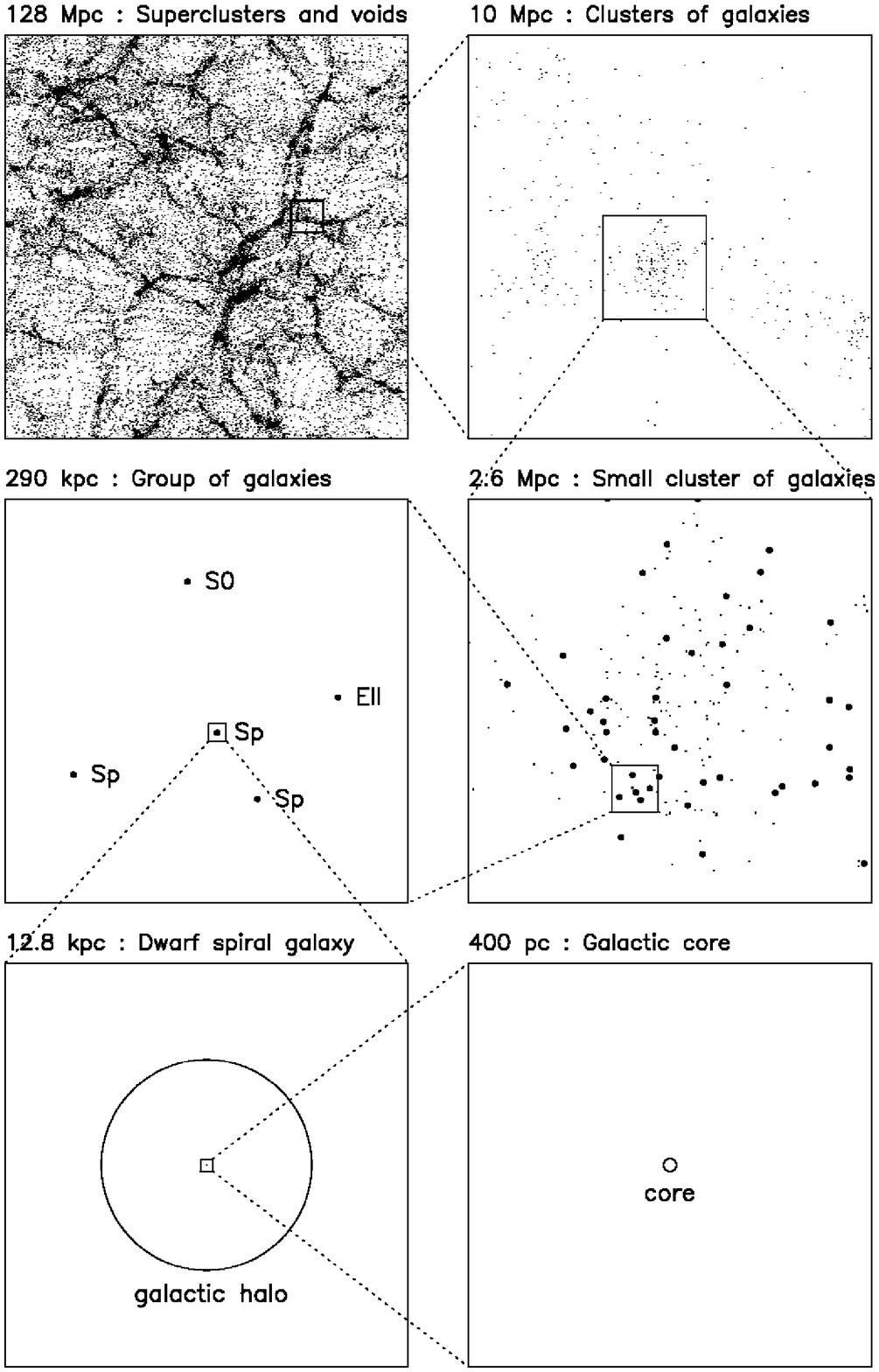}
\caption{Series of zooms illustrating the dynamical range of the
algorithm. Small dots represent $\rm P^3M$, dark matter particles.
Large dots represent actual galaxies.} 
\label{zoom}
\end{figure*}

\section{THE ALGORITHM}

We have developed a {\it multiple lens-plane algorithm} to study
light propagation in inhomogeneous universes 
\cite{pmm98,mpm00,premadietal01a,premadietal01b}. In this algorithm,
the space between
the observer and the sources is divided into a series of cubic boxes of
comoving size $128\,\rm Mpc$, and the 
matter content of each box is projected onto
a plane normal to the line of sight. The trajectories of light rays
are then computed by adding successively the deflections caused by each plane.

To use this algorithm, we need to provide a description of the matter
distribution along the line of sight. Matter is divided into two
components: background matter and galaxies. We use a $\rm P^3M$ algorithm
to simulate the distribution of background matter. The simulations used $64^3$
equal-mass particles and a $128^3$ PM grid, inside a comoving volume of
size $128\,\rm Mpc$. The matter distribution 
in the different cubes along the line of sight
then corresponds to the state of the simulation at different 
redshifts.\footnote{In practice, we combine cubes from different
simulations in order to avoid periodicities along the line of sight.}
We then use a Monte Carlo method for locating galaxies
into the computational volume \cite{mpm98,pmm98}. Galaxies are located 
according to the underlying distribution of background matter. Morphological
types are ascribed according to the morphology-density relation 
\cite{dressler80}. Galaxies are modeled as nonsingular isothermal spheres,
with rotation velocities and core radii that vary with luminosity
and morphological types. In
Figure~\ref{zoom}, we use a series of zooms to 
illustrates the dynamical range of the density distribution
generated by this method. 

\section{THE RAY-TRACING EXPERIMENTS}

We consider 3 cosmological models: (1) a flat, cosmological constant model 
with $\Omega_0=0.27$, $\lambda_0=0.73$, and $H_0=71\,\rm km/s/Mpc$.
This model is in agreement with the results of the WMAP satellite
\cite{bennettetal03}. (2) a low-density model with
$\Omega_0=0.3$, $\lambda_0=0$, and $H_0=75\,\rm km/s/Mpc$.
(3) a matter-dominated model with
$\Omega_0=1$, $\lambda_0=0$, and $H_0=75\,\rm km/s/Mpc$.
For each model, we consider sources at 5 different redshifts:
$z_s=1$, 2, 3, 4, and 5. For each combination model-redshift, we performed
10--20 ray tracing experiments. Each experiment consists of propagating a 
square beam of $101\times101=10,201$ rays back in time from the present to 
redshift $z_s$, through the matter distribution. The rays in the beam were
widely separated, by 6 arc minutes, 
and therefore sampled different regions of space.
We computed the magnification matrix ${\bf A}$ along each ray.
The magnification $\mu$ is then given by 
\begin{equation}
\mu={1\over{\rm det}\,{\bf A}}\,.
\end{equation}

\begin{figure*}[t]
\centering
\includegraphics[width=120mm]{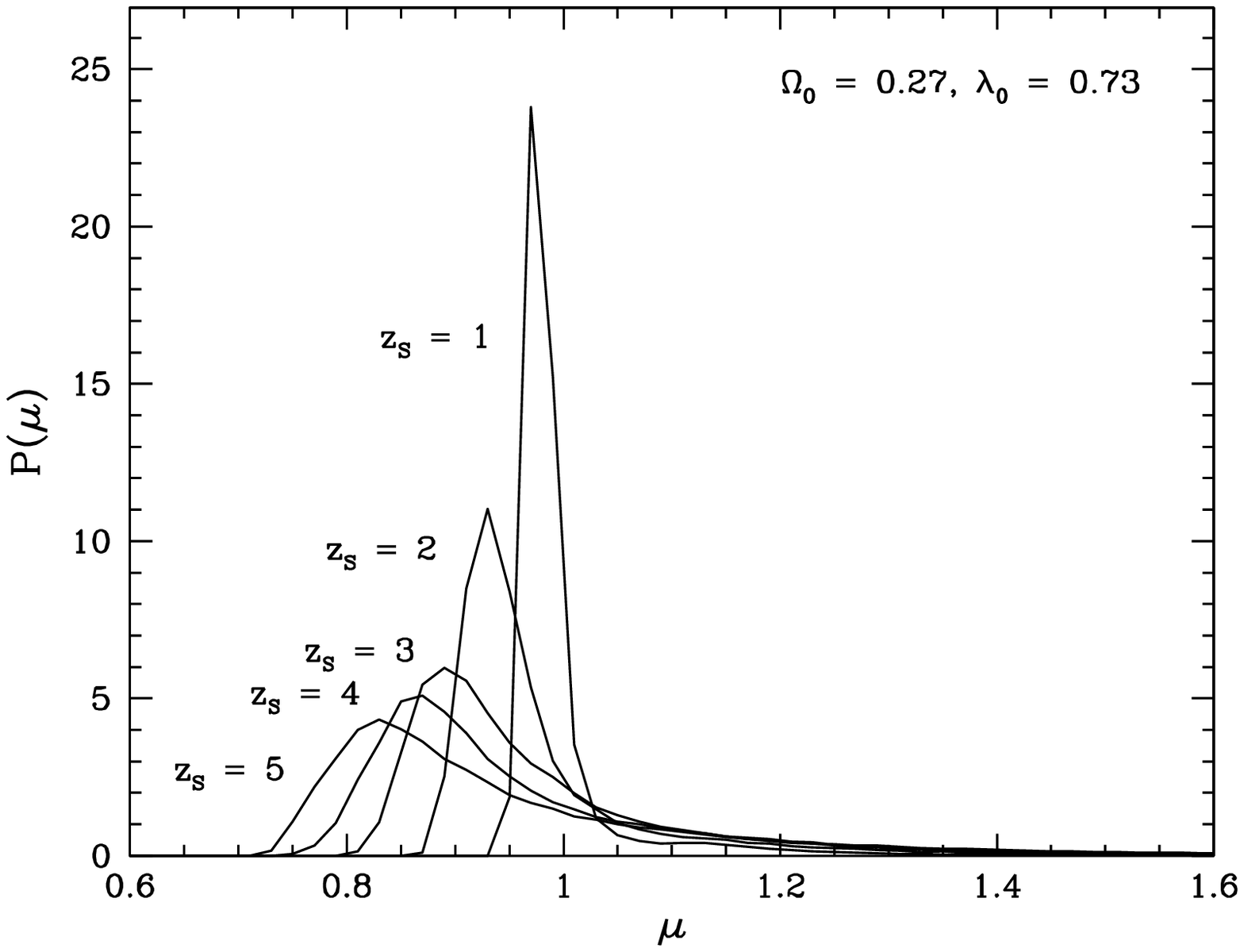}
\caption{Distribution of magnifications for the 
$\Omega_0=0.27$, $\lambda_0=0.73$ model. The various curves
correspond to different source redshifts $z_s$, as labelled.} 
\label{pmu}
\end{figure*}

\noindent
Figure~\ref{pmu} 
shows the distribution of magnifications for the $\Lambda$-model.
The distribution peaks at $\mu<1$, and is strongly skewed. The width of
the distribution increases with the source redshift. The distributions
for the other two models are qualitatively similar.

\section{THE EFFECT OF LENSING ON STATISTICS OF HIGH-Z SUPERNOVAE}

\begin{figure*}[t]
\centering
\includegraphics[width=120mm]{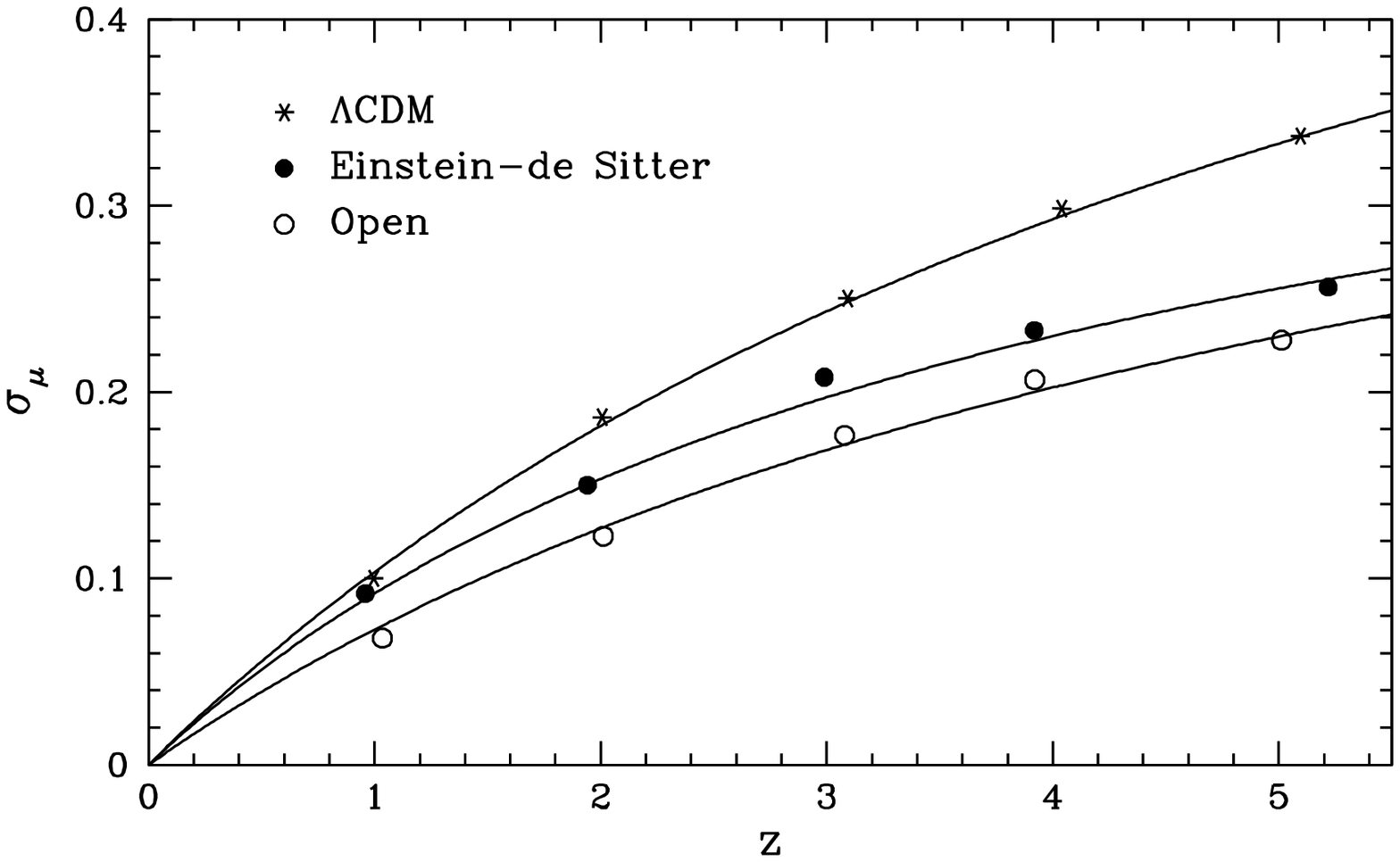}
\caption{Standard deviation $\sigma_\mu$ versus redshift, for all
three models considered. The solid lines show empirical fits of the
form $\sigma_\mu=az/(1+bz)$.}
\label{sigma}
\end{figure*}

Estimating the effect of lensing on the statistics of high-$z$ supernovae
is a complex problem, and we defer such analysis to future work. Here we
use a simple approach. First, for each model and each source redshift, we
compute the standard deviation $\sigma_\mu$ of the magnification distribution
$P(\mu)$. The values are shown in Figure~\ref{sigma}. 
We use an empirical fit of the
form 
\begin{equation}
\sigma_\mu={az\over1+bz}\,.
\end{equation} 

\noindent This enables us to estimate the values of
$\sigma_\mu$ at any redshift.

We estimate the effect of lensing as follows: the distances of high-$z$
supernovae are reported in the literature as:
\begin{equation}
\log(d_LH_0)=a\pm\delta_a\,,
\end{equation}

\noindent where $d_L$ is the luminosity distance, $H_0$ is the Hubble
constant, $a$ is the measurement, and $\delta_a$ is the intrinsic uncertainty
(i.e. not caused by lensing). The distance $d_L$ is related to the 
luminosity $L$ and flux $F$ by
\begin{equation}
F={L\over4\pi d_L^2}\,.
\end{equation}

\noindent After eliminating $d_L$, we get
\begin{equation}
(L/4\pi)^{1/2}H_0=10^a10^{\pm\delta_a}F^{1/2}\,.
\label{lsqh0}
\end{equation}

\noindent The effect of lensing will be to modify the flux $F$. To
account for it, we replace $F$ by $F\pm\Delta F$ in equation~(\ref{lsqh0}), 
and expand
to first order in $\delta_a$ and $\Delta F$. After some algebra, we get
\begin{equation}
\log(d_LH_0)=a\pm\delta_a\pm{\Delta F\over2F\ln10}\,.
\end{equation}

\begin{figure*}[t]
\centering
\includegraphics[width=120mm]{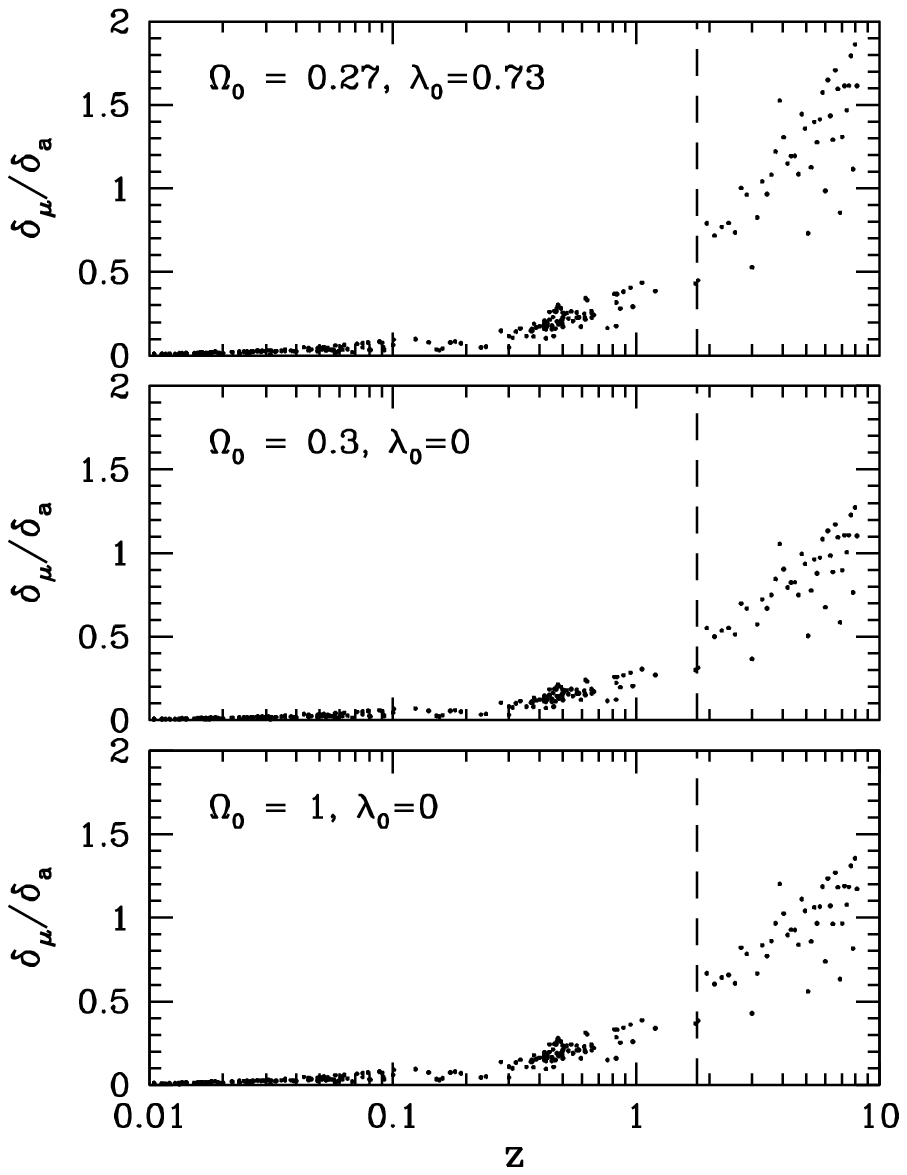}
\caption{Ratio $\delta_\mu/\delta_a$ versus redshift. The dashed
lines separate the real data of Tonry et al. (left side) from the
mock, high-redshift data (right side).}
\label{ratio}
\end{figure*}

\noindent The last term represents the effect of lensing. We then
make the approximation $\Delta F/F\approx\sigma_\mu$, and get
\begin{equation}
\log(d_LH_0)=a\pm\delta_a\pm\delta_\mu\,,
\end{equation}

\begin{figure*}[t]
\centering
\includegraphics[width=120mm]{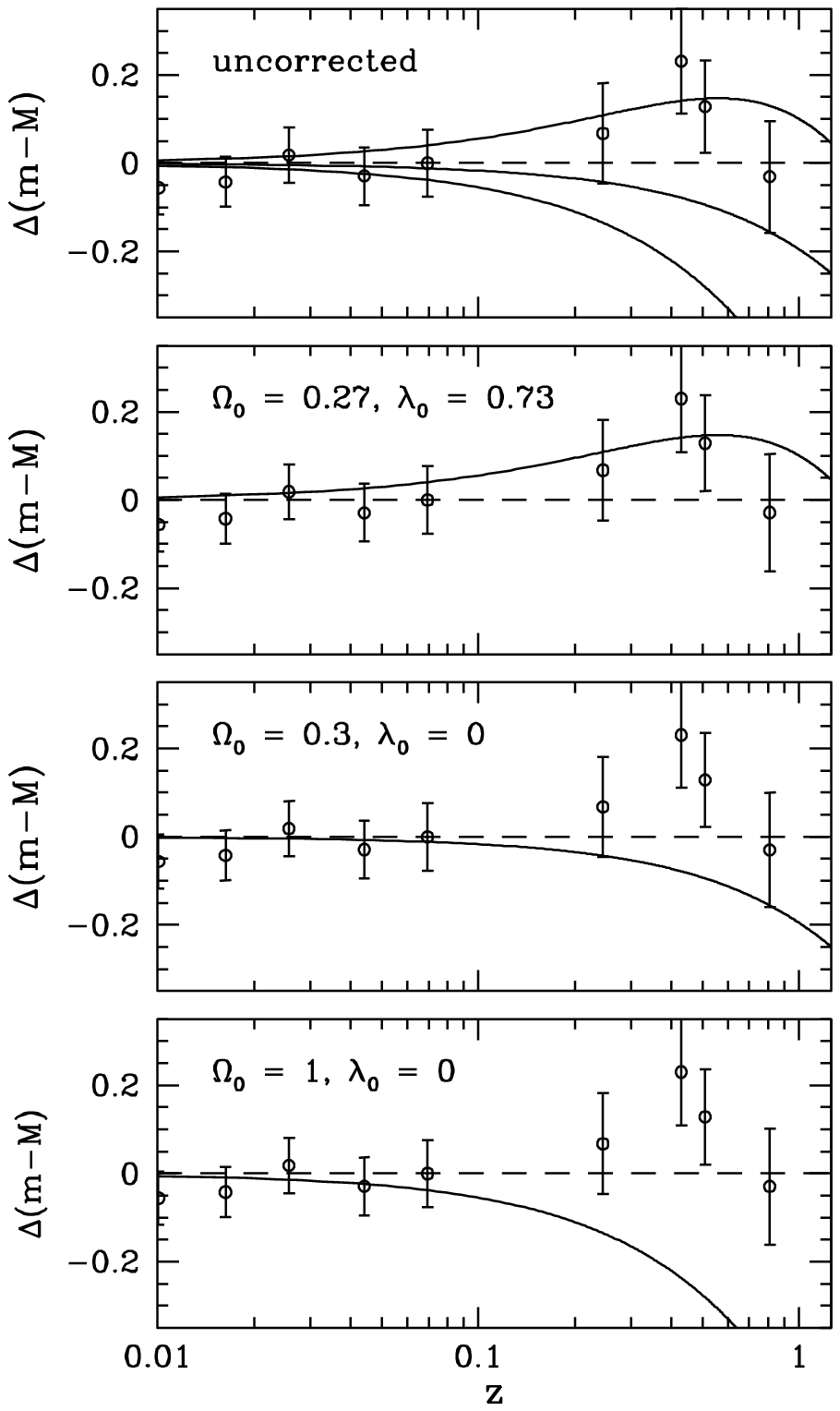}
\caption{Hubble diagram showing the magnitude deviation $\Delta(m-M)$ relative
to an empty universe, for the three models considered. In the top panel, the
three curves, from top to bottom, show the analytical result for the 
cosmological models ($\Omega_0$,$\lambda_0$)=(0.27,0.73), (0.3,0.0), 
and (1.0,0.0), respectively. The last three panels reproduce the data of the
top panel, but have been corrected to account for lensing. Since this 
correction is model-dependent, the three models are plotted on separate 
panels. Error bars show 90\% confidence level.}
\label{hubble}
\end{figure*}

\noindent where $\delta_\mu(z)=\sigma_\mu(z)/2\ln10$ is computed
using the empirical relations plotted in Figure~\ref{sigma}. We
use the values of $a$ and $\delta_a$ reported by Tonry et~al.
\cite{tonryetal03} (their Table~8). In Figure~\ref{ratio},
we plot the ratio $\delta_\mu/\delta_a$ versus $z$ (left of the dashed line).
This quantity increases with redshift, but never gets higher than 0.5 for
the Tonry et al. sample. Furthermore, we shall assume that $\delta_a$ and
$\delta_\mu$ are statistically independent, and combine them in quadrature,
using
\begin{equation}
\delta=(\delta_a^2+\delta_\mu^2)^{1/2}\,,
\end{equation}

\noindent where $\delta$ is the total error.
The contribution of lensing to this error is then of order 25\% at most.

The top panel of Figure~\ref{hubble} shows a Hubble diagram 
[deviation $\Delta(m-M)$ relative to an empty universe, versus redshift],
obtained by averaging the data in redshift bins, using
\begin{eqnarray}
w_i&=&1/\delta_i^2\,,\\
\left[\Delta(m-M)\right]_j&=&\Sigma_iw_i\Delta(m-M)/\Sigma_iw_i\,,\\
\delta_j&=&(1/\Sigma_iw_i)^{1/2}\,,
\end{eqnarray}

\noindent where the sums are over all data points $i$ in bin $j$.
The three curves, from top to bottom, show the exact results for the
$\Lambda$CDM, low-density, and matter-dominated models, respectively.
The results support the flat $\Lambda$CDM model
and exclude the other models considered.

\begin{figure*}[t]
\centering
\includegraphics[width=120mm]{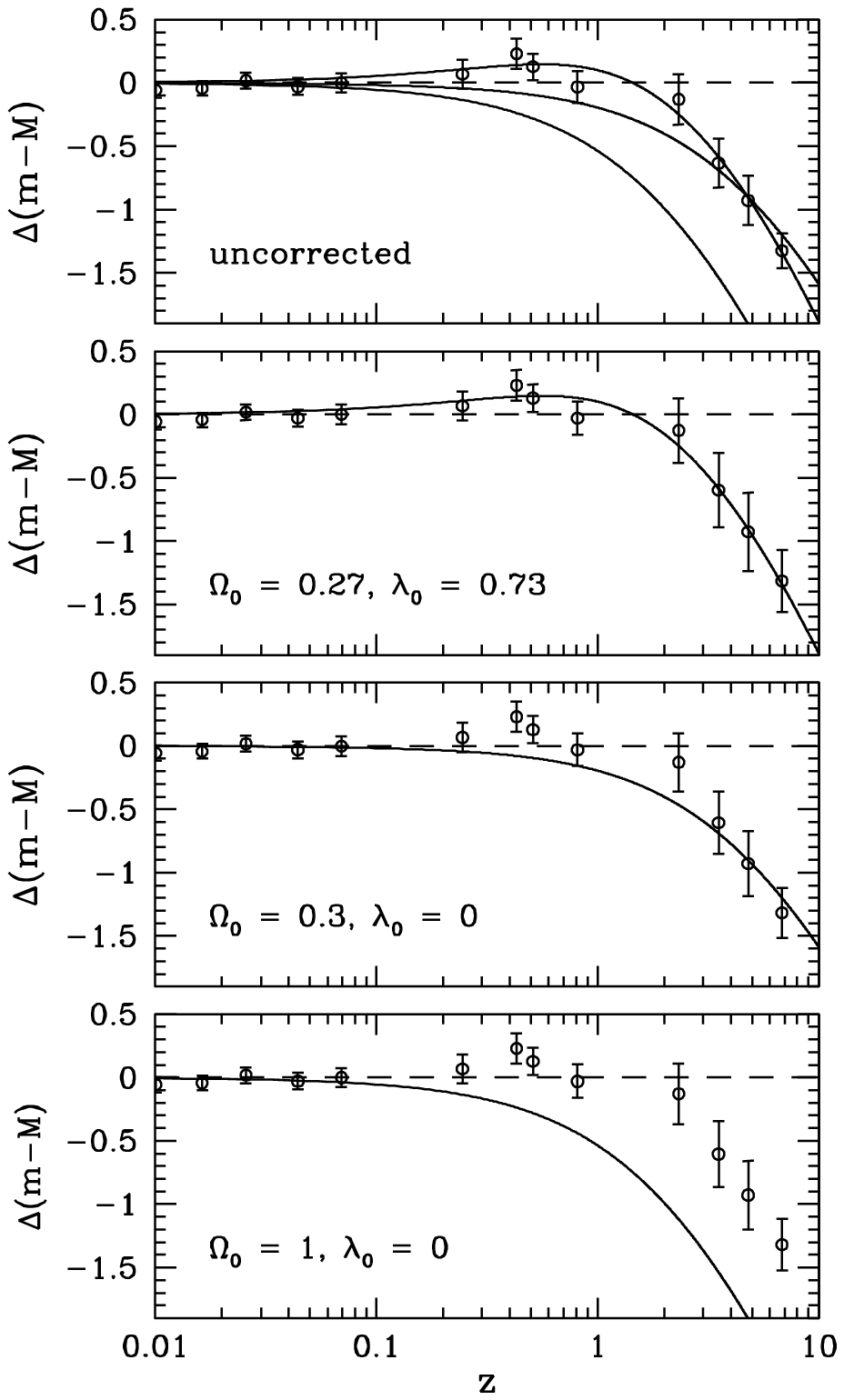}
\caption{Same as Figure \ref{hubble}, but the panels have been extended
to include the mock data (four rightmost data points).}
\label{hubble_hz}
\end{figure*}

The other panels of Figure~\ref{hubble} show the effect of lensing (the
three models have to be plotted separately, because the correction due to
lensing is model-dependent). This effect is totally negligible. The largest
correction to the error bars is about 10\% for the highest redshift bin,
for the $\Lambda$CDM model.

Clearly, the potential error introduced by lensing is negligible in 
comparison to the intrinsic error in the measurement, at least for SNe at
redshifts $z<1.8$. The next step is to estimate the effect on a 
yet-undiscovered population of very-high-$z$ SNe. We generated a mock catalog
of 43 SNe in the range $1.8<z<8$. We assume that the values of $a$ are
consistent with a $\Lambda$CDM model, and the values of $\delta_a$ in
that range are comparable to the ones in the range $1.5<z<1.8$.

The ratios $\delta_\mu/\delta_a$ are plotted in Figure~\ref{ratio}. 
The effect of lensing rapidly becomes important at redshift $z>2$. 
Figure~\ref{hubble_hz} shows the Hubble diagrams of Figure~\ref{hubble}, 
which have been extended to
higher redshifts to include the mock data. 
The error bars get significantly bigger when lensing is
included. Furthermore, at redshift $z\approx3$, it becomes very difficult
to distinguish the open, low density model from the
cosmological constant model.

\section{SUMMARY AND CONCLUSION}

We have performed a series of ray-tracing experiments using a
multiple lens-plane algorithm. We have estimated the standard deviation
of the magnification distribution with source redshift, for three
different cosmological models. Using this relation, we
have estimated the effect of lensing on the statistics of high-redshift
supernovae. The errors introduced by lensing are unimportant for SNe
with redshift $z<1.8$. However, the effect of lensing on a hypothetical
population of SNe at redshifts $z>1.8$ could be very important, and
must be understood before these SNe could be used to constrain cosmological
models.

\bigskip 
\begin{acknowledgments}
This work benefited from stimulating discussions with Dan Holtz,
Christopher Vale, and Karl Gebhardt. 
HM thanks the Canada Research Chair program and NSERC for
support.
\end{acknowledgments}

\bigskip 

\end{document}